\documentclass[smallextended]{svjour3}       % onecolumn (second format)
\smartqed  % flush right qed marks, e.g. at end of proof
\usepackage{graphicx}
\usepackage{amssymb}
\usepackage{multirow}
\journalname{Multimedia Tools and Applications}
\begin{document}

\title{Wavelet Video Coding Algorithm Based on Energy Weighted Significance Probability Balancing Tree}

\titlerunning{Wavelet Video Coding Algorithm Based on EWSPB-Tree}        % if too long for running head

\author{Chuan-Ming Song$^{1,2,3}$         \and
        Bo Fu$^{1}$   \and
        Xiang-Hai Wang$^{1,*}$  \and
        Ming-Zhe Fu$^{1}$
}

\authorrunning{C.-M. Song \and B. Fu \and X.-H. Wang \and M.-Z. Fu} % if too long for running head

%\institute{Chuan-Ming Song \at
%              School of Computer and Information Technology, Liaoning Normal University, Dalian 116029, China
%              \at
%              School of Computer Science and Technology, Dalian University of Technology, Dalian 116024, China
%              \at
%              State Key Laboratory for Novel Software Technology£¬Nanjing University, Nanjing  210093, China \\
%              Tel.: +86-411-85827711\\
%              \email{chmsong@lnnu.edu.cn}
%           Xiang-Hai Wang \at
%              School of Computer and Information Technology, Liaoning Normal University, Dalian 116029, China \\\email{xhwang@lnnu.edu.cn}
%              Bo Fu \at
%              School of Computer and Information Technology, Liaoning Normal University, Dalian 116029, China \\
%             Ming-Zhe Fu  \at
%              School of Computer and Information Technology, Liaoning Normal University, Dalian 116029, China
%}

\institute{*Corresponding author: Xiang-Hai Wang\\ \email{chmsong@163.com} \\Tel.: +86-411-85827711\\\\
$^1$ School of Computer and Information Technology, Liaoning Normal University, Dalian 116029, China \\
$^2$ School of Computer Science and Technology, Dalian University of Technology, Dalian 116024, China \\
$^3$ State Key Laboratory for Novel Software Technology, Nanjing University, Nanjing  210093, China}

\date{Received: date / Accepted: date}

\maketitle

\begin{abstract}
This work presents a 3-D wavelet video coding algorithm. By analyzing the contribution of each biorthogonal wavelet basis to reconstructed signal's energy, we weight each wavelet subband according to its basis energy. Based on distribution of weighted coefficients, we further discuss a 3-D wavelet tree structure named \textbf{significance probability balancing tree}, which places the coefficients with similar probabilities of being significant on the same layer. It is implemented by using hybrid spatial orientation tree and temporal-domain block tree. Subsequently, a novel 3-D wavelet video coding algorithm is proposed based on the energy-weighted significance probability balancing tree. Experimental results illustrate that our algorithm always achieves good reconstruction quality for different classes of video sequences. Compared with asymmetric 3-D orientation tree, the average peak signal-to-noise ratio (PSNR) gain of our algorithm are 1.24dB, 2.54dB and 2.57dB for luminance (Y) and chrominance (U,V) components, respectively. Compared with temporal-spatial orientation tree algorithm, our algorithm gains 0.38dB, 2.92dB and 2.39dB higher PSNR separately for Y, U, and V components. In addition, the proposed algorithm requires lower computation cost than those of the above two algorithms.
\end{abstract}

\section{Introduction}
\label{intro}
As the rapid development of Internet, wireless communication, and pervasive computing,
many multimedia services has been provided in various applications~\cite{01_svc_require}, such as video telephony/conferencing, mobile streaming~\cite{55_Liu}, wireless LAN video, broadband video distribution, professional video manipulation~\cite{56_Shah},
visual surveillance~\cite{50_Wu,51_Wu,52_Wu}, visual retrieval~\cite{46_Wang,49_Wang}, visual recognition~\cite{53_Wu}, visual analysis~\cite{45_Wang,47_Wang,48_Wang,54_Wu}, and etc.
These applications are able to realize cross-platform and real-time communication for clients with different power,
display resolution, and bandwidth. In these scenarios, progressive video transmission and  multi-quality services are required
due to various user requirements, client capabilities, and  transmission conditions (e.g., noise and congestion)
over heterogeneous networks.  This issues a great challenge to state-of-the-art video coding techniques, and has attracted
intensive attentions over the past decade. Scalable video coding (SVC) is
one of the effective solutions to this problem~\cite{02_overview_h264}, which encodes a video sequence once and decodes it
many times in different versions so as to efficiently adapt to the application requirements.

All the existing scalable video coding approaches can be divided into two categories. The first category represented by
MPEG-x or H.26x standards is based on closed-loop hybrid  prediction and discrete cosine transform (DCT)
structure, such as SVC amendment of H.264/MPEG-4 AVC~\cite{03_svc_amendment,svc_uhd,svc_avc} and scalable extension of HEVC~\cite{svc_extension,svc_hevc,svc_hevc_pcm,svc_hevc_low}. The second category employs
wavelet-based closed-loop~\cite{04_Marpe,05_Khan,06_zhong} or open-loop prediction structure~\cite{07_xiong,08_chen,09_vidwav_report}.
The multi-resolution property intrinsically enables wavelet transform to implement scalable video coding more easily
and flexibly than DCT. In addition, wavelet transform presents superior nonlinear approximation performance to DCT,
contributing to coding efficiency improvement. Thus, when MPEG (Moving Picture Expert Group) called for proposals
for SVC in 2003, 14 schemes were totally received worldwide, 12 of which addressed the scalable video coding
using wavelet-based approach. Moreover, studies show that closed-loop prediction structure is more efficient than open-loop
one only when the target bitrate is known, while the latter structure tends to gain superior or approximately equivalent performance to
the former in other conditions~\cite{11_lopez}. Therefore, 3-D wavelet-based scalable video coding (WSVC) exhibits great
potentials and has been widely appreciated. So far, many WSVC approaches have been proposed such as 3-D SPIHT~\cite{07_xiong},
MC-EZBC~\cite{08_chen}, VidWav platform~\cite{09_vidwav_report}, and~\cite{12_ding,13_cheng,14_chen,15_xiong,16_fang,17_tao,18_chang}.
And the Joint Video Team (JVT) originated
an ad-hoc group on ``further exploration on wavelet video coding'' in October 2004 to enhance its
coding efficiency~\cite{19_wvc_overview}. Besides all the features, e.g. spatial and temporal scalability,
provided by state-of-the-art scalable coding approaches, wavelet video coding will realize more promising functionalities~\cite{20_wvc_status}
such as very high number of spatio-temporal decomposition levels, nondyadic spatial resolution, extremely fine grain SNR scalability,
and better rate-distortion performance for very high resolution material. So far, comparison studies have demonstrated that the WSVC provides
better coding performance than those of SVC amendment of H.264/MPEG-4 AVC and the Motion JPEG-2000~\cite{svc_comparison}.
And the performance of the SVC amendment and Motion JPEG2000 depends on the resolution of coded video sequences.
However, the coding efficiency of state-of-art wavelet video coding is still slightly inferior to that of HEVC/H.265 standard.
It is thus much necessary to investigate wavelet-based video coding and further improve its rate-distortion performance.

This study presents a 3-D wavelet video coding algorithm based on energy-weighted significance probability balancing tree.
The basic idea is to weight each subband according to its wavelet basis' energy.
Subsequently, the weighted coefficients are encoded using an asymmetric tree that places the coefficients with similar
probabilities of being significant on the same layer. Our algorithm enjoys the following advantages not shared by conventional methods.

\begin{itemize}
  \item By exploiting the energy distribution of wavelet bases over different subbands, the energy-based weight can facilitate
        in theory better rate-distortion performance, as well as smaller mean squared error (MSE). Moreover, the subband weight
        raises the zerotree ratio, decreasing the synchronization information cost.
  \item The significance probability balancing tree puts those coefficients with similar probabilities of being significant on the
        same layer. Hence, the coefficients which has small probability of being significant will be moved toward leaf nodes. Through this
        way, we are able to obtain as many zerotrees as possible, and to place significant coefficients at the front of bitstream.
  \item Taking into account intra-scale relationship between neighboring coefficients, the synchronization bits are encoded block
        by block instead of the coefficient by coefficient manner of conventional algorithms. Both the synchronization information and
        computational complexity are thus efficiently reduced.
\end{itemize}

The reminder of this paper is organized as follows. Section II overviews related works. Section III discusses
the biorthogonal wavelet bases' energy of different subbands, as well as the weight of each subband. Section IV
presents a novel 3-D wavelet tree structure based on subband weight. The proposed algorithm is detailed in Section
V.  We evaluate our algorithm in Section VI and conclude the whole paper in the last section.

\section{Related Works}

Wavelet-based image and video coding has to address two key issues, namely to encode the magnitude and location
of each significant coefficient using as less bits as possible called ``synchronization information''.
For the first aspect, most algorithms employ successive approximation quantization based on bitplane to encode
significant coefficients' magnitudes, while EBCOT algorithm uses a fractional bitplane technique~\cite{21_taubman}.
These algorithms believe that the bits on the same bitplane have equal importance for the reconstructed image or video
quality regardless of whether they are in the same subband or not. In fact, this study will illustrate that the
energies of biorthogonal wavelet bases spanning different subbands generally vary. Thus, one bit cannot obtain
identical energy with that on the same bitplane but in a different subband. This indicates bitplane coding will
not achieve optimal rate-distortion performance. For the second aspect, EZW~\cite{23_Shapiro}, SPIHT~\cite{24_Said},
and SLCCA~\cite{25_Chai} employ zerotree to locate significant coefficients according to magnitude attenuation from
coarse to fine scales. In contrast, SPECK~\cite{26_Pearlman}, EZBC~\cite{27_Hsiang}, and EBCOT use block, called zeroblock,
as a unit to transmit synchronization information exploiting intra-scale correlation of wavelet coefficients.

Moinuddin \emph{et al.} pointed out that the tree structure plays a significant role in improving 3-D wavelet video
coding efficiency~\cite{32_Moinuddin,33_Fowler}. Motivated by the zerotree and zeroblock in image compression,
researchers extended symmetrically these two 2-D
structures to 3-D cases for video coding~\cite{07_xiong,08_chen,28_Campisi,29_Vass,30_Kim,31_Xu,34_Chen,35_Khalil,36_Minami},
e.g. \cite{28_Campisi}, \cite{29_Vass}, \cite{07_xiong} and \cite{30_Kim}, \cite{31_Xu}, and \cite{08_chen}
are separately the extensions of EZW, SLCCA, SPIHT, EBCOT, and EZBC. Nevertheless, the wavelet coefficient
distribution of still images is obviously different from that of video frames, especially temporal high-pass
frames. \cite{37_He} calculates the average standard deviation (STD) of Carphone, Mother \& Daughter, and Hall Monitor sequences
along horizontal, vertical, and temporal direction. Statistics show that the STD along temporal direction is much smaller
than the STDs along the other two directions, while the STDs along horizontal and vertical direction are very close.
In this case, the amplitudes of temporal high-pass coefficients tend to be smaller than those of spatial
high-pass coefficients. The probabilities of the coefficients on the same layer being significant are nonuniform
under such a symmetric tree structure. As a result, part of the synchronization information is wasted, inevitably affecting
the overall coding efficiency. In general, instead of intuitively extending zerotree or
zeroblock from 2-D to 3-D case, the distribution characteristics of wavelet coefficients should be taken into
account when designing tree structure. This will definitely reduce further the overhead of synchronization
information and achieve better rate-distortion performance.

Kim \emph{et al.} use an approximate symmetry tree structure in~\cite{07_xiong} and obtain superior coding efficiency
to symmetry tree structure. However, the approximate symmetry tree requires that the numbers of wavelet decomposition
along spatial and temporal directions be equal. This restriction cannot fully exploit the redundancy along temporal
direction. To address this issue, an asymmetric 3-D orientation tree is proposed in~\cite{37_He} which attaches all
subbands together to form a longer subband tree. This asymmetric 3-D orientation tree has no limitation on the number
of wavelet decomposition along each direction and outperforms 3-D SPIHT. Compared with symmetric tree structure, the
asymmetric structure in~\cite{37_He} is capable of gathering more insignificant coefficients, requiring less synchronization
bits. In order to build a longer tree, a virtual zerotree~\cite{12_ding,38_Khan} is proposed as an extension of
existing tree structures. It virtually creates zerotrees in $LL$ subband so that the significant map can be coded
in a more efficient way, although no decimation and decomposition actually takes place. However,~\cite{12_ding,37_He,38_Khan}
is still ignored the nonuniform probabilities of spatio-temporal coefficients being significant on the same layer.
Considering this fact, Zhang \emph{et al.}~\cite{39_Zhang} decomposes a spatio-temporal orientation tree into a
temporal direction tree and a spatial orientation tree.
Moreover, the temporal direction trees are encoded with a higher priority over spatial orientation trees.
Only when a temporal direction tree has a significant coefficient, the spatial orientation tree which
the significant temporal coefficient belongs to will be scanned. This method however may delay the coding of isolated zeros
in spatial orientation trees.

We have discussed the inter-scale correlation of wavelet coefficients above. In fact there is another correlation,
i.e. intra-scale correlation. For the video sequence following a stable random process, the intra-scale correlation
is even stronger than inter-scale correlation~\cite{40_Song}. On the basis of asymmetric spatio-temporal orientation
tree, \cite{32_Moinuddin,41_Moinuddin} introduce the zeroblock structure of size $2 \times 2$ pixels into 3-D video
coding and code the synchronization information using zeroblock as a unit. Since this method takes advantage of
correlations of both inter- and intra-scale, its video coding efficiency is effectively improved. But \cite{32_Moinuddin,41_Moinuddin}
treat the coefficients in different subbands indiscriminately. In the next section, we will discuss the unequal roles
of different subbands in terms of energy for the reconstructed image/video quality.

\section{Energy-Based Weight of Wavelet Coefficients }

Most state-of-the-art wavelet-based video coders employ biorthogonal wavelet transform and encode significant
coefficients bitplane by bitplane. These approaches believe the bits on the same bitplane in different subbands are of equal
importance. However,~\cite{42_Usevitch,43_Usevitch} points out that biorthogonal wavelet transform does not
have energy-preserving property. Consequently, the coefficients in separate subbands work differently for the
reconstructed signal energy.

%Thus, it is unreasonable to treat coefficients of all subbands in the same way.
%In this section, we will prove the reconstructed energy of wavelet coefficients of each subband is different with biorthogonal wavelet transform and propose a coefficient weighting method based reconstructed energy.

Let $g$ and $h$ be the low-pass and high-pass filter of biorthogonal wavelet, whose length are $M$ and
$N$, respectively. Their dual filters are separately  $\tilde{g}$ and $\tilde{h}$. And the low-frequency
and high-frequency coefficients under scale $j$ are denoted by $a_i^j$ and $b_i^j$, in which $0 \leq i<L$ and $L$ is the
length of low-frequency and high-frequency components. Then the  $1$-level inverse wavelet transform
can be expressed as follows,
\begin{equation}\label{inverse_wavelet}
    a^{j + 1}  = \tilde g * \left( {Ua^j } \right) + \tilde h * \left( {Ub^j } \right),
\end{equation}
where  $U$ and ``*'' are up-sampling and convolution operators, respectively. Because wavelet transform $f$
is linear, we have $f(x)=xf(1)$. Thus we only need to examine the  energy of unit coefficient in
each subband to prove that the bases of separate subbands contribute differently to reconstructed image/video.

%it requires to prove the different reconstructed energy of each subband coefficients in units because of the both only differ one factor  $x^2$.

%\newtheorem{theorem}{Proof}
\begin{theorem} [Wavelet bases' energy]
Let  $a^j_K=1$ ($K$ is a constant, $\left\lfloor {N/2} \right\rfloor  \le K \le L - \left\lfloor {N/2} \right\rfloor$),
while the other coefficients in $a^j$ and all coefficients in $b^j$ be 0. Suppose the reconstructed energy after
inverse transform is $E_a$. Likewise, set  $b_K^j=1$ ($K$ is a constant, $\left\lfloor {M/2} \right\rfloor  \le K \le L - \left\lfloor {M/2} \right\rfloor$), while the other coefficients in $b^j$ and all coefficients in $a^j$ to be 0. Let the
energy after inverse transform now be $E_b$. Then we have $E_a \neq E_b$.
\end{theorem}

%\begin{IEEEproof}
    \textbf{Proof}: In the case of $a^{j + 1}  = \tilde g * \left( {Ua^j } \right)$, from the procedure of wavelet transform,
    a nonzero value will produce only if the filter coefficients coincide with $a_K^j$. Then we have $E_a  = \sum\nolimits_{l = 0}^{N - 1} {\left( {\tilde g_l } \right)^2 } $, where $\tilde g_l $ stands for the $l$th filter coefficient of $\tilde g$.
    Similarly, we can obtain $E_b  = \sum\nolimits_{l = 0}^{M - 1} {\left( {\tilde h_l } \right)^2 } $ in the case of $a^{j + 1}  = \tilde h * \left( {Ub^j } \right)$. For all commonly used biorthogonal wavelets in image and video compression, e.g. Daubechies 9/7 and 5/3 wavelet,
    their sums of squared filter coefficients are not equal. Therefore, we conclude that $E_a  \ne E_b $.
%\end{IEEEproof}

Theorem 1 indicates that we cannot achieve the minimized MSE if we employ traditional bitplane
technique to encode all subbands' coefficients. We thus propose to weight the wavelet coefficients subband by subband
before bitplane coding. The weight of each subband equals the square root of its unit coefficient's energy. Note that
the subband weight varies with its corresponding wavelet basis. Table~\ref{subband_weight} lists the weights of 160
temporal-spatio subbands obtained by a 4-level 5/3 temporal decomposition followed by a 3-level 9/7 spatial decomposition,
in which the line and column represent temporal and spatial subbands, respectively. From Table~\ref{subband_weight} we
can find that the lower the subband, the larger the weight. Moreover, there exist obvious differences among the weights
of different subbands. This result not only verifies Theorem 1, but also illustrates the necessity of weighting wavelet
coefficients.

\begin{table*}[!t]
\renewcommand{\arraystretch}{1.3}
\caption{Weights of Spatio-temporal Subbands After a 4-level Temporal Decomposition Followed by a 3-level Spatial Decomposition}
\label{subband_weight}
\centering
\begin{tabular}{ccccccccccc}
\hline
\hline
    & $LL_3$ & $LH_3$ &	$HL_3$ & $HH_3$ &	$LH_2$ &	$HL_2$ &	$HH_2$ &	$LH_1$ &	$HL_1$ &	$HH_1$ \\
\hline
$LLLL$  &9.71&7.31&7.31&5.50&5.28&5.28&3.87&4.04&4.04&3.15\\
$LLLH$  &4.03&3.04&3.04&2.28&2.19&2.19&1.61&1.68&1.68&1.31\\
$LLH_1$ &2.98&2.24&2.24&1.69&1.62&1.62&1.19&1.24&1.24&0.97\\
$LLH_2$ &3.98&2.99&2.99&2.25&2.16&2.16&1.58&1.66&1.66&1.29\\
$LH_1$  &2.17&1.63&1.63&1.22&1.18&1.18&0.87&0.90&0.90&0.70\\
$LH_2$  &2.33&1.75&1.75&1.32&1.27&1.27&0.93&0.97&0.97&0.75\\
$LH_3$  &2.49&1.88&1.88&1.41&1.36&1.36&0.99&1.04&1.04&0.81\\
$LH_4$  &2.77&2.09&2.09&1.57&1.51&1.51&1.10&1.15&1.15&0.90\\
$H_1$   &2.06&1.55&1.55&1.17&1.12&1.12&0.82&0.86&0.86&0.67\\
$H_2$   &2.06&1.55&1.55&1.17&1.12&1.12&0.82&0.86&0.86&0.67\\
$H_3$   &2.06&1.55&1.55&1.17&1.12&1.12&0.82&0.86&0.86&0.67\\
$H_4$   &2.06&1.55&1.55&1.17&1.12&1.12&0.82&0.86&0.86&0.67\\
$H_5$   &2.06&1.55&1.55&1.17&1.12&1.12&0.82&0.86&0.86&0.67\\
$H_6$   &2.06&1.55&1.55&1.17&1.12&1.12&0.82&0.86&0.86&0.67\\
$H_7$   &2.13&1.60&1.60&1.20&1.15&1.15&0.85&0.85&0.88&0.67\\
$H_8$   &1.94&1.46&1.46&1.10&1.06&1.06&0.77&0.81&0.81&0.67\\
\hline
\hline
\end{tabular}
\end{table*}

\section{3-D Significance Probability Balancing Tree}

After weighting each subband, an appropriate tree structure is needed to better locate significant coefficients. Most
previous asymmetrical structures pursuit tree depth without considering the nodes' importance distribution on
the same tree level. Table~\ref{average_energy_comparison} shows average energy comparison between spatial and temporal
high-frequency coefficients of the first 16 frames of ``Foreman'' and ``Mobile \& Calendar'' sequences after 3-D wavelet transform.
It can be seen that the average energy of different high-frequency subbands vary apparently, especially $LLLL$
and $LLLH$ subbands having the largest energy. The average amplitude of spatial high-frequency coefficients is larger than that of temporal high-frequency
coefficients. This indicates that the probability of spatial high-frequency coefficients being significant
is higher than that of temporal high-frequency coefficients. If both of them are placed on the same level,
the latter has to be tested repeatedly before being significant so that the synchronization bits will be wasted.
To address this issue, we present a novel 3-D tree structure named \emph{significance probability balancing tree} in
this section.

\begin{table*}[!t]
\renewcommand{\arraystretch}{1.0}
\caption{Average Energy Comparison Between Spatial and Temporal High-frequency Coefficients of Foreman and Mobile \& Calendar
         Sequences After 3-D Wavelet Transform}
\label{average_energy_comparison}
\centering
\begin{tabular}{ccccc}
\hline
\hline
\multirow{3}{*}{\textbf{Temporal}} &  \multicolumn{2}{c}{\textbf{Foreman}} &	\multicolumn{2}{c}{\textbf{Mobile \& Calendar}} \\ \cline{2-3} \cline{4-5}
                      &  \textbf{Spatial}	&  \textbf{Temporal}	& \textbf{Spatial}	& \textbf{Temporal} \\
   \textbf{Subbands}  &  \textbf{Subbands'}	&  \textbf{Subbands'}	& \textbf{Subbands'}	& \textbf{Subbands'} \\
                      &  \textbf{Energy}	&  \textbf{Energy}	& \textbf{Energy}	& \textbf{Energy} \\
\hline
$LLLL$  &39779&6005&119298&5308\\
$LLLH$  &1133&2172&2344&2056\\
$LLH_0$ &319&199&390&62\\
$LLH_1$ &757&515&1158&247\\
$LH_0$  &53&36&56&20\\
$LH_1$  &62&36&88&21\\
$LH_2$  &76&32&123&26\\
$LH_3$  &202&75&239&33\\
\hline
\hline
\end{tabular}
\end{table*}

\begin{figure*}[!t]
\centering
    \includegraphics[width=4.8in]{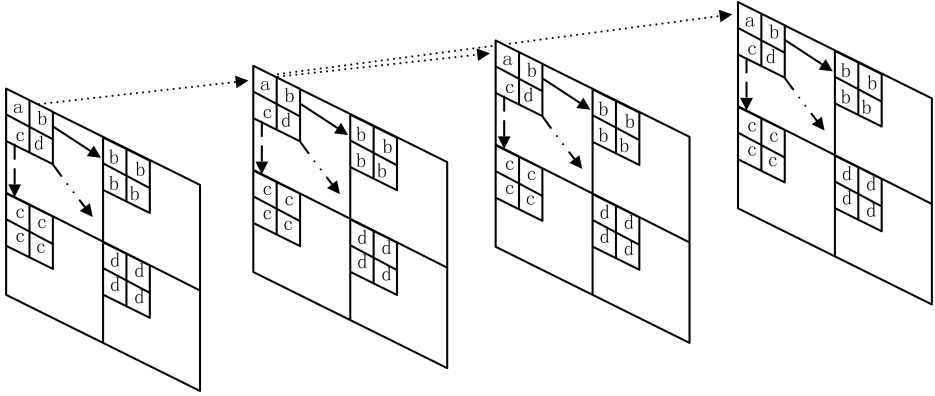}
    \caption{Parent-child relationship of the proposed 3-D significance probability balancing tree}
    \label{fig1}
\end{figure*}

Our basic idea is to place those coefficients with similar probabilities of being significant on the same layer
based on the amplitude correlation of spatio-temporal coefficients. To construct a significance probability balancing tree,
one approach is to study the coefficients' distribution before each scan, and then to establish an adaptive
tree structure. But the high computational demand will inhibit its practical use. Our proposed method processes
the descendants of a parent node along its spatial and temporal orientation, respectively. The spatial descendants
are arranged using spatial orientation tree of SPIHT, while the temporal descendants are organized from coarse
to fine scales along temporal direction, selecting the spatial node with no offsprings as a root.
Fig.~\ref{fig1} illustrates the parent-child relationship of the proposed tree. For the sake of clarity, only four temporal frames
are shown with a $2$-level temporal decomposition followed by a $1$-level spatial decomposition. %But trees grow along
%the temporal direction in the same fashion as Fig.~\ref{fig1} in all but the temporal highest frequency frames.

Furthermore, adjacent wavelet coefficients in a subband tend to have identical importance~\cite{40_Song}. We thus group adjacent
$2 \times 2$ coefficients into a block in the temporal high-frequency frames. Each block is treated as
an offspring of the coefficient at the corresponding position in temporal low-frequency subband.  The structure
discussed above is named ``\emph{temporal-domain block tree}''. Only if a block contains the root(s) of nonzero subtree(s),
the block will be divided into four coefficients, one of which is the root of a temporal-domain block tree. The
other coefficients are separately the roots of three spatial orientation trees. Therefore, the proposed
significance probability balancing tree is essentially a hybrid spatial orientation tree and temporal-domain
block tree structure.

\begin{figure}[!t]
\centering
    \includegraphics[width=3.0in]{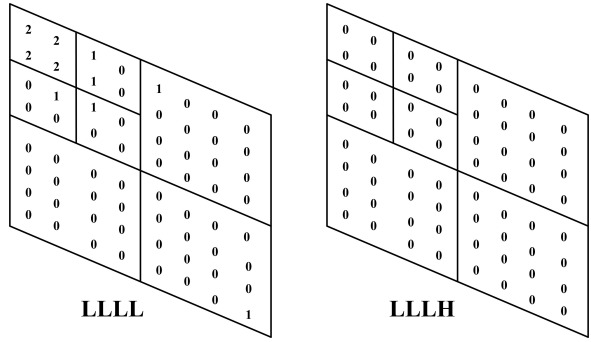}
    \caption{Part of quantized coefficients of temporal $LLLL$ and $LLLH$ frames of ``Foreman''}
    \label{fig2}
\end{figure}

When using our proposed tree to organize 3-D wavelet coefficients, the synchronization information is
coded block by block instead of coefficient by coefficient. Hence, this block-wise manner is able to reduce
effectively the number of synchronization bits. In order to verify this point, we decompose the first 16 frames
of ``Foreman'', ``Hall Monitor'', and ``Mobile \& Calendar'' by 3-D wavelet transform and represent the resulting coefficients
using our proposed tree and a typical asymmetric tree~\cite{37_He}. Subsequently, we calculate the ratio
of degree-1 zerotree and degree-2 zerotree~\cite{44_Cho} under the above two tree structures, respectively.
As depicted in Table~\ref{zerotree ratio}, the proposed significance probability balancing tree can achieve
higher zerotree ratio compared with the asymmetrical tree~\cite{37_He}. Fig.~\ref{fig2} presents part of  quantized
coefficients of temporal $LLLL$ and $LLLH$ frames of ``Foreman'', where ``2'' , ``1'', and ``0''  separately denote
coded coefficients, significant coefficients, and insignificant coefficients. And the quantization
step size is 256. Suppose that the coding of sign bits is ignored. Then the asymmetrical tree ~\cite{37_He}
needs a total of 40 bits namely ``1101010100110001000010011100000000100010'' to code significance map, while
the significance probability balancing tree requires only 35 bits, i.e. ``0110101010011 00010111000000000100 01''.
For the other two sequences, a similar results can be obtained. Note that the above three test sequences belong
to different classes of videos in MPEG-4 test library. ``Foreman'' has low spatial detail and medium amount of movement,
while ``Mobile \& Calendar'' has high spatial detail and medium amount of movement, and ``Hall Monitor'' contains
low spatial detail and low amount of movement. Consequently, We can conclude that our proposed tree  consumes
less synchronization bits than typical asymmetric tree in general cases.

\begin{table*}[!t]
\renewcommand{\arraystretch}{1.3}
\caption{The Zerotree Ratio Between Our Tree Structure and Asymmetric Tree Structure (\%)}
\setlength{\tabcolsep}{0.8mm}
\label{zerotree ratio}
\centering
\begin{tabular}{ccccccc}
\hline
\hline
\multirow{3}{*}{\textbf{Scan No.}} &  \multicolumn{2}{c}{\textbf{Foreman}} &	\multicolumn{2}{c}{\textbf{Hall Monitor}} &  \multicolumn{2}{c}{\textbf{Mobile \& Calendar}} \\ \cline{2-7}
                  &  \textbf{Asymmetric}	& \textbf{Proposed}	& \textbf{Asymmetric}	& \textbf{Proposed}	& \textbf{Asymmetric}	 & \textbf{Proposed} \\
                  & \textbf{Tree}	& \textbf{Tree}	& \textbf{Tree}	& \textbf{Tree}	& \textbf{Tree}	 & \textbf{Tree} \\
\hline
$1^{st}$ \textbf{Scan} &100.00&100.00&100.00&100.00&100.00&100.00\\
$2^{nd}$ \textbf{Scan} &100.00&100.00&99.79&100.00&98.26&100.00\\
$3^{rd}$ \textbf{Scan} &90.91&98.96&88.96&99.92&70.74&92.63\\
$4^{th}$ \textbf{Scan} &77.12&92.72&76.39&97.44&59.15&69.17\\
$5^{th}$ \textbf{Scan} &63.62&80.79&70.06&87.52&53.85&61.31\\
$6^{th}$ \textbf{Scan} &59.20&67.19&62.81&77.99&49.86&57.23\\
$7^{th}$ \textbf{Scan} &55.94&63.35&56.38&71.60&41.40&56.18\\
\hline
\hline
\end{tabular}
\end{table*}

\section{Implementation of Proposed 3-D Wavelet Video Coding Algorithm}

Based on the weighted coefficients and significance probability balancing tree, this section
presents a novel 3-D wavelet video coding algorithm.

Similar to SPIHT algorithm, we employ three ordered lists to store the coefficients' significance
information, namely list of insignificant sets (LIS), list of insignificant pixels (LIP), and list
of significant pixels (LSP). To facilitate process of the descendants of node $(x,y)$, we use $D(x,y)$
and $L(x,y)$ to denote separately the coordinate set of all descendants of $(x,y)$ and the
coordinate set of all descendants except all offsprings of $(x,y)$. Furthermore, an element $(x,y)$
in LIS is either $D(x,y)$ or $L(x,y)$. To differentiate between them, we name this element ``TYPE\_A entry'' if
it represents $D(x,y)$, while ``TYPE\_B entry'' if it represents $L(x,y)$. Assuming $T_i$ and
$T_0$ denotes quantization threshold of the $i$th scan and the initial threshold respectively,
we detail the implementation of our proposed coding algorithm as below.

%It is composed of
%four processes, namely 3-D wavelet transform, initialization, search significant coefficients,
%and significant coefficients refinement of amplitude.

\begin{enumerate}
    \item[\textbf{1.}]  Parse the input video into a number of GOPs, and then apply 3-D wavelet transform to each GOP.
    \item[\textbf{2.}]  \textbf{Initialization}.
    \begin{enumerate}
        \item [2.1] Calculate the initial threshold $T_0$ for each GOP as follows
                    \begin{equation} \label{find_threshold}
                        {T_0} = \left\lfloor {{{\log }_2}\left( {\mathop {\max }\limits_{{C_n} \in C} \left| {{c_n}} \right|} \right)} \right\rfloor,
                    \end{equation}
                    where $C$ represents the 3-D wavelet coefficient set.
        \item [2.2] Set $i=0$ and LSP $={\O}$. Add the coefficients of lowpass subband in the temporally lowest-frequency frame,
       e.g. the upper-left ``a'', ``b'', ``c'' and ``d'' of the first frame as depicted in Fig.~\ref{fig1}, to LIP and LIS, and set them in LIS as TYPE\_A entries.
    \end{enumerate}
    \item[\textbf{3.}] \textbf{Search for significant coefficients.}
        \begin{enumerate}
            \item[3.1] Compare each coefficient $(x,y) \in$ LIP with $T_i$. If $|(x,y)| \geq T_i$, output ``1'' and its sign bit, and then move $(x,y)$ to LSP. Otherwise, output ``0''.
            \item[3.2] For each untreated $(x,y) \in$ LIS, if it is a node of spatial orientation tree, e.g. ``b'', ``c'', and ``d'' of the first frame in Fig.~\ref{fig1}, go to Step 3.3. Otherwise, if it is a node of temporal-domain block tree, e.g. ``a'' of the first frame in Fig.~\ref{fig1}, go to Step 3.4.
            \item[3.3] Code $(x,y)$ using SPIHT, go to Step 3.5.
            \item[3.4] If $(x,y)$ is a TYPE\_B entry, go to Step 3.4.4. Else, if $D(x,y)$ does not contain significant coefficients, output ``0'' and go to Step 3.5. Otherwise, output ``1''.
            \begin{enumerate}
              \item[3.4.1] Check whether there are significant coefficients in $(x,y)$'s temporal-domain child blocks, e.g. the upper-left ``a'', ``b'', ``c'', and ``d'' of the second frame depicted in Fig.~\ref{fig1}. If only insignificant coefficients are contained, output ``0'' and move $(x,y)$ into the LIP. Go to Step 3.4.3.
              \item[3.4.2] Otherwise, output ``1'' and test each node in every child block of $(x,y)$. For each significant coefficient, output ``1'' and sign bit, move it into LSP. For insignificant coefficient, output ``0'' and move it to LIP.
              \item[3.4.3] If $L(x,y)=\phi$, remove $(x,y)$ from LIS; Otherwise, move $(x,y)$ to the end of the LIS as an entry of TYPE\_B and go to Step 3.5.
              \item[3.4.4] If $L(x,y)$ does not contain significant coefficients, output ``0''. Otherwise, output ``1'', and split current block into one root of temporal-domain block subtree and three roots of spatial directional tree. Move the four roots into LIS as entries of TYPE\_A and remove $(x,y)$ from LIS.
            \end{enumerate}
            \item[3.5] If all $(x,y) \in$ LIS have been coded, go to Step 4. Otherwise, go to Step 3.2.
        \end{enumerate}
        \item[\textbf{4.}] \textbf{Refine the amplitudes of significant coefficients.}
        \begin{enumerate}
          \item [4.1] For each entry $(x,y) \in$ LSP, if $(x,y) \in [T_i,1.5T_i)$, output ``0''. Otherwise, output ``1''.
          \item [4.2] If target rate has been reached, return. Otherwise, Set $i=i+1$, $t_i=T_{i-1}/2$.
          \item [4.3] If $T_i=0$, return. Else, go to Step 3.
        \end{enumerate}
\end{enumerate}

\section{Experimental Results and Analysis}

Extensive experiments were conducted on seven color video sequences including the first 128 frames of ``Foreman'', ``Hall Monitor'',
``Mobile \& Calendar'', ``Coastguard'', ``Mother \& Daughter'', ``Miss America'', and ``Bus'' in CIF@ 30Hz format.
All experiments were performed on VidWav platform~\cite{09_vidwav_report}. The 3-D wavelet transform was decomposed in ``t+2D'' manner
with $4$-level 5/3 motion-compensated temporal filtering and $3$-level 2-D 9/7 wavelet. Motion estimation were carried out with quarter-pixel accuracy. For each color video sequence, its Y, U, and V components were sequentially encoded.

To verify the effectiveness of our algorithm, we compare it against two representative methods, i.e. the asymmetric 3-D orientation tree \cite{37_He} and temporal-spatial orientation tree~\cite{39_Zhang}, in terms of peak signal-to-noise ratio (PSNR).
Note that, the PSNR statistics of \cite{37_He} were obtained on VidWav platform, while the results of temporal-spatial orientation
tree were extracted from~\cite{39_Zhang} which only presented the PSNRs for ``Foreman'', ``Miss America'', and ``Mobile \& Calendar''.
Table~\ref{coding_results} shows the PSNR comparison result among the above three coding algorithms at 128Kbps-1500Kbps for
``Miss America'', ``Foreman'', and ``Mobile \& Calendar''. Table~\ref{coding_results_hall}-Table~\ref{coding_results_bus}
list the comparison results between the asymmetric 3-D orientation tree and our proposed algorithm for ``Hall Monitor'', ``Mother \& Daughter'', ``Coastguard'', and ``Bus'' at 128Kbps-1500Kbps.

As can be seen from Table~\ref{coding_results}-Table~\ref{coding_results_bus},
asymmetric 3-D orientation tree outperforms temporal-spatial orientation tree for those sequences with low amount
of movement, such as Miss America and Foreman,  while the latter gains superior
efficiency to the former for sequences with high spatial detail and medium amount of movement,
such as Mobile \& Calendar. This indicates that these two tree structures have distinct merits and their performances
always depend on the characteristics of video sequences. Since our algorithm takes into consideration the coefficients' significance
probability distribution on the same tree level, it is less sensitive to video characteristics and
achieves the highest PSNR for all test sequences.
For Y component, the average PSNR of proposed algorithm is separately 1.24dB and 0.38dB higher than those of \cite{37_He} and
\cite{39_Zhang}. While for U and V components, our algorithm gains 2.54dB and 2.57dB higher PSNR compared with \cite{37_He},
and 2.92dB and 2.39dB higher PSNR than \cite{39_Zhang}. It is worth mentioning that for the Y component of ``Mobile \& Calendar'',
the PSNR achieved by our algorithm is lower than that of ~\cite{39_Zhang} as shown in Table~\ref{coding_results}.
According to the experimental results
presented in~\cite{39_Zhang}, the PSNR improvement of \cite{39_Zhang} is separately 0.18dB, 0.34dB, and 0.41dB at 500Kbps,
1000Kbps, and 1500Kbps compared with \cite{37_He}. Nevertheless, our algorithm obtains 1.92dB, 1.26dB, and 1.49dB
higher PSNR than those of \cite{37_He} at the above bitrates, respectively. In this sense, our algorithm still outperforms
\cite{39_Zhang} regarding the Y component of ``Mobile \& Calendar''.

\begin{table*}[!t]
\renewcommand{\arraystretch}{1.3}
\caption{PSNR Comparison Among Three Test Algorithms at Different Bitrates}
\setlength{\tabcolsep}{0.8mm}
\label{coding_results}
\centering
\begin{tabular}{ccccccccccc}
\hline
\hline
\multirow{3}{*}{\textbf{Test Sequence}} & \raisebox{-1.5ex}[0pt][0pt]{\textbf{Bitrate}} &  \multicolumn{9}{c}{\textbf{PSNR (dB)}} \\ \cline{3-11}
    & \raisebox{-1.5ex}[0pt][0pt]{\textbf{(Kbps)}}  & \multicolumn{3}{c}{\textbf{Asymmetric Tree}} & \multicolumn{3}{c}{\textbf{Temporal-spatial Tree}} & \multicolumn{3}{c}{\textbf{Proposed Tree}} \\ \cline{3-11}
    &   & Y & U & V & Y & U & V& Y & U & V \\
\hline
\multirow{7}{*}{\textbf{Miss America}} & 128&39.99&38.35&40.41&-----&-----&-----&40.88&39.10&41.99\\
                                       & 256&40.57&38.73&41.01&-----&-----&-----&42.38&39.92&43.35\\
&384&41.91&39.84&42.14&-----&-----&-----&43.05&40.45&43.95\\
&500&42.34&40.23&42.54&41.76&40.64&42.27&43.38&40.83&44.17\\
&768&42.78&40.68&43.11&-----&-----&-----&44.04&41.84&44.68\\
&1000&42.88&40.84&43.23&43.23&40.69&43.46&44.36&42.51&45.00\\
&1500&43.96&42.32&44.42&43.99&40.69&43.46&44.84&43.82&45.52\\
\hline
\multirow{7}{*}{\textbf{Foreman}} & 128&29.04&35.45&35.41&-----&-----&-----&29.80&37.53&37.39\\
&256&32.30&37.87&38.38&-----&-----&-----&33.32&39.79&40.43\\
&384&33.79&38.70&39.86&-----&-----&-----&35.07&40.90&42.10\\
&500&35.04&39.52&40.97&34.84&37.63&39.42&36.18&41.71&43.08\\
&768&36.37&40.49&42.20&-----&-----&-----&37.81&43.12&44.51\\
&1000&37.54&41.26&43.26&37.76&39.63&41.76&38.96&44.01&45.37\\
&1500&38.99&42.72&44.71&39.40&40.50&42.62&40.51&45.42&46.66\\
\hline
\multirow{7}{*}{\textbf{Mobile \& Calendar}} & 128&20.38&25.99&24.75&-----&-----&-----&20.76&27.87&26.38\\
&256&23.15&28.28&27.19&-----&-----&-----&23.99&30.31&29.06\\
&384&24.63&29.21&28.45&-----&-----&-----&25.71&31.84&30.67\\
&500&25.05&29.52&28.83&27.38&30.92&30.97&26.97&33.54&32.44\\
&768&27.55&31.28&30.92&-----&-----&-----&28.57&35.19&34.35\\
&1000&28.56&32.28&32.07&31.26&34.38&34.55&29.82&36.82&36.01\\
&1500&30.00&33.70&33.36&33.49&35.77&35.99&31.49&38.47&37.76\\
\hline
\hline
\end{tabular}
\end{table*}

\begin{table*}[!t]
\renewcommand{\arraystretch}{1.3}
\caption{PSNR Comparison Between Asymmetric 3-D Orientation Tree Algorithm and Our Algorithm at Different Bitrates for ``Hall Monitor" sequence}
\label{coding_results_hall}
\centering
\begin{tabular}{ccccccc}
\hline
\hline
%&  \multicolumn{6}{c}{\textbf{PSNR (dB)}} \\ \cline{2-7}
%& \multicolumn{6}{c}{\textbf{Hall Monitor}} \\ \cline{2-7}
\textbf{Bitrate}& \multicolumn{3}{c}{\textbf{Asymmetric 3-D Tree}} & \multicolumn{3}{c}{\textbf{Proposed Algorithm}} \\ \cline{2-7}
\textbf{(Kbps)}    & Y & U & V & Y & U & V \\
\hline
128&31.01&36.47&38.88&31.96&37.68&40.21   \\
256&34.64&37.57&39.72&36.46&39.28&41.46   \\
384&36.71&38.42&40.57&38.10&39.89&42.14    \\
500&37.32&38.72&40.82&38.96&40.32&42.48   \\
768&38.54&39.77&41.57&40.14&41.16&43.03   \\
1000&39.22&40.69&42.08&40.74&41.71&43.43 \\
1500&39.68&41.43&42.52&41.70&42.66&44.16 \\
\hline \hline
\end{tabular}
\end{table*}

\begin{table*}[!t]
\renewcommand{\arraystretch}{1.3}
\caption{PSNR Comparison Between Asymmetric 3-D Orientation Tree Algorithm and Our Algorithm at Different Bitrates for ``Mother \& Daughter" sequence}
\label{coding_results_mother}
\centering
\begin{tabular}{ccccccc}
\hline
\hline
\textbf{Bitrate} & \multicolumn{3}{c}{\textbf{Asymmetric 3-D Tree}} & \multicolumn{3}{c}{\textbf{Proposed Algorithm}} \\ \cline{2-7}
\textbf{(Kbps)}    & Y & U & V & Y & U & V \\
\hline
128&35.20&40.52&41.18&36.25&42.90&43.40   \\
256&38.57&42.40&43.54&39.55&44.85&45.73   \\
384&40.27&43.56&44.71&41.19&45.97&46.67    \\
500&41.36&44.45&45.48&42.36&46.64&47.42   \\
768&42.48&45.50&46.20&43.83&47.56&48.42   \\
1000&43.59&46.26&47.00&44.70&48.10&48.98 \\
1500&44.56&47.14&47.91&45.62&48.64&49.52 \\
\hline \hline
\end{tabular}
\end{table*}

%\begin{table*}[!t]
%\renewcommand{\arraystretch}{1.3}
%\caption{PSNR Comparison Between Asymmetric 3-D Orientation Tree Algorithm and Our Algorithm at Different Bitrates}
%\label{coding_results_hall_mother}
%\centering
%\begin{tabular}{ccccccc|cccccc}
%\hline
%\hline
%&  \multicolumn{12}{c}{\textbf{PSNR (dB)}} \\ \cline{2-13}
%\textbf{Bitrate}    & \multicolumn{6}{c}{\textbf{Hall Monitor}} & \multicolumn{6}{c}{\textbf{Mother \& Daughter}} \\ \cline{2-13}
%\textbf{(Kbps)}    & \multicolumn{3}{c}{\textbf{Asymmetric 3-D Tree}} & \multicolumn{3}{c}{\textbf{Proposed Algorithm}} & \multicolumn{3}{c}{\textbf{Asymmetric 3-D Tree}} & \multicolumn{3}{c}{\textbf{Proposed Algorithm}} \\ \cline{2-13}
%    & Y & U & V & Y & U & V& Y & U & V & Y & U & V \\
%\hline
%128&31.01&36.47&38.88&31.96&37.68&40.21&35.20&40.52&41.18&36.25&42.90&43.4   \\
%256&34.64&37.57&39.72&36.46&39.28&41.46&38.57&42.40&43.54&39.55&44.85&45.73   \\ 384&36.71&38.42&40.57&38.10&39.89&42.14&40.27&43.56&44.71&41.19&45.97&46.67    \\
%500&37.32&38.72&40.82&38.96&40.32&42.48&41.36&44.45&45.48&42.36&46.64&47.42   \\ 768&38.54&39.77&41.57&40.14&41.16&43.03&42.48&45.50&46.20&43.83&47.56&48.42   \\ 1000&39.22&40.69&42.08&40.74&41.71&43.43&43.59&46.26&47.00&44.70&48.10&48.98 \\
%1500&39.68&41.43&42.52&41.70&42.66&44.16&44.56&47.14&47.91&45.62&48.64&49.52 \\
%\hline \hline
%\end{tabular}
%\end{table*}

\begin{table*}[!t]
\renewcommand{\arraystretch}{1.3}
\caption{PSNR Comparison Between Asymmetric 3-D Orientation Tree Algorithm and Our Algorithm at Different Bitrates for ``Coastguard" sequence}
\label{coding_results_coast}
\centering
\begin{tabular}{ccccccc}
\hline
\hline
\textbf{Bitrate}& \multicolumn{3}{c}{\textbf{Asymmetric 3-D Tree}} & \multicolumn{3}{c}{\textbf{Proposed Algorithm}} \\ \cline{2-7}
\textbf{(Kbps)}  & Y & U & V & Y & U & V \\
\hline
128&24.96&37.63&36.24&25.58&40.40&41.66\\
256&26.35&38.40&37.51&27.61&41.63&43.32\\
384&27.94&40.06&41.06&28.89&42.45&44.14\\
500&28.67&40.29&41.16&29.80&42.99&44.54\\
768&29.88&40.89&42.09&31.29&43.64&45.18\\
1000&31.32&41.38&43.14&32.52&44.22&45.73\\
1500&33.10&42.10&43.66&34.30&45.02&46.48\\
\hline \hline
\end{tabular}
\end{table*}

\begin{table*}[!t]
\renewcommand{\arraystretch}{1.3}
\caption{PSNR Comparison Between Asymmetric 3-D Orientation Tree Algorithm and Our Algorithm at Different Bitrates for ``Bus" sequence}
\label{coding_results_bus}
\centering
\begin{tabular}{ccccccc}
\hline
\hline
\textbf{Bitrate}& \multicolumn{3}{c}{\textbf{Asymmetric 3-D Tree}} & \multicolumn{3}{c}{\textbf{Proposed Algorithm}} \\ \cline{2-7}
\textbf{(Kbps)}  & Y & U & V & Y & U & V \\
\hline
128&21.04&31.49&29.87&21.43&34.56&35.67\\
256&24.42&34.40&34.95&25.14&36.49&37.79\\
384&25.84&34.99&35.71&27.01&37.36&38.87\\
500&27.28&35.94&37.13&28.09&37.82&39.45\\
768&29.13&36.62&38.03&30.37&39.16&40.92\\
1000&30.10&37.30&38.76&31.48&39.69&41.44\\
1500&32.41&38.30&39.99&34.03&41.37&43.01\\
\hline \hline
\end{tabular}
\end{table*}

In addition, the ratio of zerotrees is increased after the wavelet coefficients are weighted, as illustrated in Table~\ref{zerotree ratio}.
Thus, the number of isolated zeroes is effectively reduced, which always involve many comparisons and input/output
operations in conventional zerotree coding algorithms. Further, since the proposed temporal-domain block tree operates on
units of size $2 \times 2$ pixels, the nunber of entries in LIP and LIS is less than that of \cite{37_He} and \cite{39_Zhang}.
On one hand, the energy-based weight and temporal-domain block tree are capable of improving the efficiency of synchronization information. On the
other hand, they help to lower the computational complexity of video coding algorithm. Of course, with the increase of target bitrate,
the temporal-domain block trees need to be recursively split, the computational complexity will gradually
close to that of \cite{37_He,39_Zhang}.

\section{Conclusions}

In this study, by analyzing the contribution of each biorthogonal wavelet basis in terms of its reconstructed energy,
we proposed to weight each subband by the energy of its corresponding basis before encoding.
According to the distribution of weighted coefficients,
we put forward a concept of 3-D significance probability balancing tree structure and implement it using  hybrid spatial
orientation tree and temporal-domain block tree. Consequently, a novel 3-D wavelet video coding algorithm is presented
based on energy-weighted significance probability balancing tree. We verify its effectiveness through extensive experiments.
We believe that our study will be certainly useful in future researches and developments of wavelet-based scalable video
coding.

\begin{acknowledgements}
This work is supported by the National Natural Science Foundation of China (NSFC) under Grant nos. 61402214,
41671439, and 61702246, and the Open Foundation of State Key Laboratory for Novel Software Technology of Nanjing
University under Grant no. KFKT2018B07, and the Dalian Foundation for Youth Science and Technology Star (2015R069).
\end{acknowledgements}

%\bliographystyle{spphys}       % APS-like style for physics

\bibliographystyle{elsarticle-num}
%\bibliography{IEEEabrv,latex}
\bibliography{latex}

\begin{thebibliography}{10}
\expandafter\ifx\csname url\endcsname\relax
  \def\url#1{\texttt{#1}}\fi
\expandafter\ifx\csname urlprefix\endcsname\relax\def\urlprefix{URL }\fi
\expandafter\ifx\csname href\endcsname\relax
  \def\href#1#2{#2} \def\path#1{#1}\fi

\bibitem{01_svc_require}
{SVC} requirements specified by {MPEG, JVT-N026}, Tech. rep., {ISO/IEC
  JTC1/SC29/WG11}, Hong Kong (2005).

\bibitem{55_Liu}
Y.~Liu, J.~Y.~B. Lee, Post-streaming rate analysis¡ªa new approach to mobile
  video streaming with predictable performance, IEEE Transactions on Mobile
  Computing 16~(12) (2017) 3488--3501.

\bibitem{56_Shah}
R.~Shah, P.~J. Narayanan, Interactive video manipulation using object
  trajectories and scene backgrounds, IEEE Transactions on Circuits and Systems
  for Video Technology 23~(9) (2013) 1565--1576.

\bibitem{50_Wu}
L.~Wu, Y.~Wang, J.~Gao, X.~Li, Deep adaptive feature embedding with local
  sample distributions for person re-identification, Pattern Recognition 73~(1)
  (2018) 275--288.

\bibitem{51_Wu}
L.~Wu, Y.~Wang, X.~Li, J.~Gao, What-and-where to match: Deep spatially
  multiplicative integration networks for person re-identification, Pattern
  Recognition 76~(1) (2018) 727--738.

\bibitem{52_Wu}
L.~Wu, Y.~Wang, Z.~Ge, Q.~Hu, X.~Li, Structured deep hashing with convolutional
  neural networks for fast person re-identification, Computer Vision and Image
  Undersanding 167~(1) (2018) 63--73.

\bibitem{46_Wang}
Y.~Wang, X.~Lin, L.~Wu, W.~Zhang, Effective multi-query expansions:
  Collborative deep networks for robust landmark retrieval, IEEE Transactions
  on Image Processing 26~(3) (2017) 1393--1404.

\bibitem{49_Wang}
Y.~Wang, L.~Wu, Beyond low-rank representations: Orthogonal clustering basis
  reconstruction with optimized graph structure for multi-view spectral
  clustering, Neural Networks 103~(1) (2018) 1--8.

\bibitem{53_Wu}
L.~Wu, Y.~Wang, X.~Li, J.~Gao, Deep attention-based spatially recursive
  networks for fine-grained visual recognition, IEEE Transactions on
  Cybernetics 99~(PP).
\newblock \href {http://dx.doi.org/10.1109/TCYB.2018.2813971}
  {\path{doi:10.1109/TCYB.2018.2813971}}.

\bibitem{45_Wang}
Y.~Wang, X.~Lin, L.~Wu, W.~Zhang, Q.~Zhang, X.~Huang, Robust subspace
  clustering for multi-view data by exploiting correlation consensus, IEEE
  Transactions on Image Processing 24~(11) (2015) 3939--3949.

\bibitem{47_Wang}
Y.~Wang, W.~Zhang, L.~Wu, X.~Lin, X.~Zhao, Unsupervised metric fusion over
  multiview data by graph random walk-based cross-view diffusion, IEEE
  Transactions on Neural Networks and Learning Systems 28~(1) (2017) 57--70.

\bibitem{48_Wang}
Y.~Wang, W.~Zhang, L.~Wu, X.~Lin, M.~Fang, S.~Pan, Iterative views agreement:
  An iterative low-rank based structured optimization method to multi-view
  spectral clustering, in: Proc. International Joint Conference on Artificial
  Intelligence, Vol.~1, New York, USA, 2016, pp. 2153--2159.

\bibitem{54_Wu}
Y.~Wang, L.~Wu, X.~Lin, J.~Gao, Multiview spectral clustering via structured
  low-rank matrix factorization, IEEE Transactions on Neural Networks and
  Learning Systems 99~(PP).
\newblock \href {http://dx.doi.org/10.1109/TNNLS.2017.2777489}
  {\path{doi:10.1109/TNNLS.2017.2777489}}.

\bibitem{02_overview_h264}
H.~Schwarz, D.~Marpe, T.~Wiegand, Overview of the scalable video coding
  extension of the {H.264/AVC} standard, IEEE Transactions on Circuits and
  Systems for Video Technology 17~(9) (2007) 1103--1120.

\bibitem{03_svc_amendment}
T.~Wiegand, G.~J. Sullivan, J.~Reichel, H.~Schwarz, H.~Wien, Joint draft 11 of
  {SVC} amendment, {Doc. JVT-X201}, Tech. rep., Geneva, Switzerland (2007).

\bibitem{svc_uhd}
U.-K. Park, H.~Choi, J.~W. Kang, J.-G. Kim, Scalable video coding with large
  block for {UHD} video, IEEE Transactions on Consumer Electronics 28~(3)
  (2012) 932--940.

\bibitem{svc_avc}
A.~Bjelopera, S.~Grgic, Scalable video coding extension of {H.264/AVC}, in:
  Proc. {International Symposium ELMAR}, Vol.~1, Zadar, Croatia, 2012, pp.
  7--12.

\bibitem{svc_extension}
{ISO/IEC JTC 1/SC 29/WG 11 and ITU-T SG 16 WP 3}, Joint call for proposals on
  scalable video coding extensions of high efficiency video coding {(HEVC)},
  Tech. Rep. {N12957}, Stockholm, Sweden (2012).

\bibitem{svc_hevc}
Z.~Shi, X.~Sun, F.~Wu, Spatially scalable video coding for {HEVC}, IEEE
  Transactions on Circuits and Systems for Video Technology 22~(12) (2012)
  1813--1826.

\bibitem{svc_hevc_pcm}
G.~Wu, W.~Ding, Y.~Shi, B.~Yin, Adaptive weighted prediction for scalable video
  coding based on {HEVC}, in: Proc. {Pacific Rim Conference on Multimedia
  (PCM)}, Vol.~1, London, 2013, pp. 110--121.

\bibitem{svc_hevc_low}
S.~Lasserre, F.~L. Leannec, J.~Taquet, E.~Nassor, Low-complexity intra coding
  for scalable extension of {HEVC} based on content statistics, IEEE
  Transactions on Circuits and Systems for Video Technology 24 (2014) (to be
  published).

\bibitem{04_Marpe}
D.~Marpe, H.~L. Cycon, Very low bit-rate video coding using wavelet-based
  techniques, IEEE Transactions on Circuits and Systems for Video Technology
  9~(1) (1999) 85--94.

\bibitem{05_Khan}
E.~Khan, M.~Ghanbari, An efficient and scalable low bit-rate video coding with
  virtual {SPIHT}, Signal Processing: Image Communication 19~(3) (2004)
  267--283.

\bibitem{06_zhong}
M.~S. Zhong, M.~Ghanbari, Motion compensation based on wavelet coefficient
  blocks, Acta Automatica Sinica 30~(1) (2004) 64--69.

\bibitem{07_xiong}
B.~Kim, Z.~Xiong, W.~A. Pearlman, Low bit rate, scalable video coding with
  {3-D} set partitioning in hierarchical trees {(3-D SPIHT)}, IEEE Transactions
  on Circuits and Systems for Video Technology 10~(12) (2000) 1374--1387.

\bibitem{08_chen}
P.~S. Chen, J.~W. Woods, Bidirectional {MC-EZBC} with lifting implementation,
  IEEE Transactions on Circuits and Systems for Video Technology 14~(10) (2004)
  1183--1194.

\bibitem{09_vidwav_report}
{ISO/IEC JTC1/SC29/WG11}, Wavelet codec reference document and software manual,
  Tech. Rep. {ISO/MPEG Video, Tech. Rep. N7334} (Jul. 2005).

\bibitem{11_lopez}
M.~F. L\`{o}pez, V.~G. Ruiz, I.~Garc\'{\i}a, Efficiency of closed and open-loop
  scalable wavelet based video coding, Advanced Concepts for Intelligent Vision
  Systems, Lecture Notes in Computer Science 4678~(10) (2007) 800--809.

\bibitem{12_ding}
W.~Ding, J.~Hu, L.~Zhang, Optimal {3D-SPIHT} video coding method by reducing
  redundancy between trees, Journal of Computer-Aided Design \& Computer
  Graphics 17~(3) (2005) 563--569.

\bibitem{13_cheng}
C.~C. Cheng, G.~J. Peng, W.~L. Hwang, Subband weighting with pixel connectivity
  for {3-D} wavelet coding, IEEE Transactions on Image Processing 18~(1) (2009)
  52--62.

\bibitem{14_chen}
P.~Chen, J.~W. Woods, Improved {MC-EZBC} with quarter-pixel motion vectors,
  Tech. rep., {ISO/IEC JTC 1/SC 29/WG 11}, Fairfax, VA.

\bibitem{15_xiong}
R.~Xiong, J.~Xu, F.~Wu, S.~Li, Barbell-lifting based {3-D} wavelet coding
  scheme, IEEE Transactions on Circuits and Systems for Video Technology 17~(9)
  (2007) 1256--1269.

\bibitem{16_fang}
S.~Fang, Y.~Zhong, {3D} subband codec with full scalability, Mini-Micro Systems
  26~(7) (2005) 1260--1263.

\bibitem{17_tao}
T.~J, H.~Wang, J.~Zhang, Z.~Jiang, Research on the scalability of {3D} wavelet
  video coding, Mini-Micro Systems 26~(2) (2005) 285--288.

\bibitem{18_chang}
Z.~Chang, L.~Zhuo, L.~Shen, An improved motion-compensated three dimension
  wavelet video coding method, Journal of Circuits and Systems 11~(1) (2006)
  113--117, 121.

\bibitem{19_wvc_overview}
{ISO/IEC JTC 1/SC 29/WG 11}, Wavelet video coding - an overview, {Doc. W7824},
  Bangkok, Thailand (2006).

\bibitem{20_wvc_status}
{ISO/IEC JTC 1/SC 29/WG 11}, Status report - version 1 on wavelet video coding
  exploration, {Doc. N7822}, Bangkok, Thailand (2006).

\bibitem{svc_comparison}
X.~Lu, G.~R. Martin, Performance comparison of the {SVC, WSVC, and Motion
  JPEG2000} advanced scalable video coding schemes, in: Proc. {IET, Intelligent
  Signal Processing}, Vol.~1, London, 2013, pp. 1--6.

\bibitem{21_taubman}
D.~Taubman, High performance scalable image compression with {EBCOT}, IEEE
  Transactions on Image Processing 9~(7) (2000) 1158--1170.

\bibitem{23_Shapiro}
J.~Shapiro, Embedded image coding using zerotree of wavelet coefficients, IEEE
  Transactions on Signal Processing 41~(12) (1993) 3445--3462.

\bibitem{24_Said}
A.~Said, W.~A. Pearlman, A new, fast, and efficient image codec based on set
  partitioning in hierarchical trees, IEEE Transactions on Circuits and Systems
  for Video Technology 6~(3) (1996) 243--250.

\bibitem{25_Chai}
B.~Chai, J.~Vass, X.~Zhuang, Significance link connected component analysis for
  wavelet image coding, IEEE Transactions on Image Processing 8~(6) (1999)
  774--784.

\bibitem{26_Pearlman}
W.~A. Pearlman, A.~Islam, N.~Nagaraj, A.~Said, Efficient low-complexity image
  coding with set-partitioning embedded block coder, IEEE Transactions on
  Circuits and Systems for Video Technology 14~(11) (2004) 1219--1235.

\bibitem{27_Hsiang}
S.~T. Hsiang, J.~W. Woods, Embedded image coding using zeroblocks of
  subband/wavelet coefficient and context modeling, in: Proc. {IEEE}
  {International Symposium on Circuit and Systems (ISCAS'00)}, Vol.~3, Geneva,
  Switzerland, 2000, pp. 662--665.

\bibitem{32_Moinuddin}
A.~A. Moinuddin, E.~K. E, M.~Ghanbari, The impact of tree structures on the
  performance of zerotree-based wavelet video codecs, Signal Processing: Image
  Communication 25~(3) (2010) 179--195.

\bibitem{33_Fowler}
J.~E. Fowler, B.~Pesquet-Popescu, An overview on wavelets in source coding,
  communications, and networks, EURASIP Journal on Image and Video Processing
  2007~(1) (2007) 1--27.

\bibitem{28_Campisi}
P.~Campisi, M.~Gentile, A.~Neri, Three-dimensional wavelet-based approach for a
  scalable video conference system, in: Proc. {IEEE} {International Conference
  on Image Processing (ICIP'99)}, Vol.~3, Kobe, Japan, 1999, pp. 802--806.

\bibitem{29_Vass}
J.~Vass, B.~Chai, X.~Zhuang, {3-D SLCCA}---a highly scalable very low bit-rate
  software-only wavelet video codec, in: Proc. {IEEE Second Workshop Multimedia
  Signal Processing}, Vol.~3, Redondo Beach, CA, 1998, pp. 474--479.

\bibitem{30_Kim}
B.~J. Kim, W.~A. Pearlman, An embedded wavelet video coder using
  three-dimensional set partitioning in hierarchical trees ({SPIHT}), in: Proc.
  Data Compression Conference, Vol.~1, Snowbird, USA, 1997, pp. 251--260.

\bibitem{31_Xu}
J.~Xu, Z.~Xiong, S.~Li, Y.-Q. Zhang, Three-dimensional embedded subband coding
  with optimal truncation ({3-D ESCOT}), Applied Computational Harmonic
  Analysis 10~(3) (2001) 290--315.

\bibitem{34_Chen}
Y.~Chen, W.~A. Pearlman, Three-dimensional subband coding of video using the
  zerotree method, in: Proc. {SPIE Visual Communications and Image Processing},
  Vol. 2727, Snowbird, USA, 1996, pp. 1302--1312.

\bibitem{35_Khalil}
H.~Khalil, F.~A. F, S.~I. Shaheen, Lowering frame-buffering requirements of
  {3-D} wavelet transform coding of interactive video, in: Proc. {IEEE}
  {International Conference on Image Processing (ICIP'99)}, Vol.~3, Kobe,
  Japan, 1999, pp. 852--856.

\bibitem{36_Minami}
G.~Minami, Z.~Xiong, A.~Wang, P.~A. Chou, S.~Mehrotra, {3-D} wavelet coding of
  video with arbitrary regions of support, IEEE Transactions on Circuits and
  Systems for Video Technology 11~(9) (2001) 1063--1068.

\bibitem{37_He}
C.~He, J.~Dong, Y.~Zheng, Z.~Gao, Optimal {3-D coefficient tree structure for
  3-D} wavelet video coding, IEEE Transactions on Circuits and Systems for
  Video Technology 13~(10) (2003) 961--972.

\bibitem{38_Khan}
E.~Khan, M.~Ghanbari, Very low bit rate video coding using virtual spiht,
  Electronic Letters 37~(1) (2001) 40--41.

\bibitem{39_Zhang}
L.~Zhang, D.~Wang, A.~Vincent, Decoupled {3-D} zerotree structure for
  wavelet-based video coding, IEEE Transactions on Broadcasting 54~(3) (2008)
  430--436.

\bibitem{40_Song}
C.-M. Song, X.-H. Wang, F.~Zhang, Visually lossless accuracy of motion vector
  in overcomplete wavelet-based scalable video coding, Journal of Computers
  4~(9) (2009) 821--828.

\bibitem{41_Moinuddin}
A.~A. Moinuddin, E.~Khan, M.~Ghanbari, Efficient and embedded {3-D} wavelet
  video coding, in: Proc. {TENCON 2008}, Vol.~1, Hyderabad, 2008, pp. 1--4.

\bibitem{42_Usevitch}
B.~Usevitch, Optimal bit allocation for biorthogonal wavelet coding, in: Proc.
  Data Compression Conference, Vol.~1, Snowbird, USA, 1996, pp. 387--395.

\bibitem{43_Usevitch}
B.~E. Usevitch, A tutorial on modern lossy wavelet image compression:
  foundations of {JPEG 2000}, IEEE Signal Processing Magazine 18~(5) (2001)
  22--35.

\bibitem{44_Cho}
Y.~Cho, W.~A. Pearlman, Quantifying the coding performance of zerotrees of
  wavelet coefficients: degree-{k} zerotree, IEEE Transactions on Signal
  Processing 55~(6) (2007) 2425--2431.

\end{thebibliography}

%\begin{thebibliography}{}
%%
%% and use \bibitem to create references. Consult the Instructions
%% for authors for reference list style.
%\end{thebibliography}

\end{document}